# Stripes of Partially Fluorinated Alkyl Chains: Dipolar Langmuir Monolayers

Matthias F. Schneider[1,‡,*], David Andelman[1,2], Motomu Tanaka[1, †*]

[1] Lehrstuhl für Biophysik, Technische Universität München, 85748 Garching, Germany

[2] School of Physics and Astronomy, Raymond and Beverly Sackler Faculty of Exact Sciences, Tel Aviv University, Ramat Aviv 69978, Tel Aviv, Israel

[‡] Present address: Lehrstuhl für Experimentalphysik I, Biophysics, Universität Augsburg, Universitätsstraße 1, D-86135 Augsburg, Germany

[†] On leave at Department of Physics and Center of Excellence, Kyoto University, Japan

* Corresponding authors:

Matthias F. Schneider

Lehrstuhl für Experimentalphysik I, Biophysics, Universität Augsburg, Universitätsstraße 1,

D-86135 Augsburg, Germany

Tel: ++49-821-598-3311, Fax: ++49-821-5983225, e-mail: matthias.schneider@physik.uni-augsburg.de

Motomu Tanaka

Lehrstuhl für Biophysik, Technische Universität München, 85748 Garching, Germany

Tel: ++49-89-2891-2539, Fax: ++49-89-2891-2469, e-mail: mtanaka@ph.tum.de

**Abstract**

Stripe-like domains of Langmuir monolayers formed by surfactants with partially fluorinated lipid anchors (F-alkyl lipids) are observed at the gas-liquid phase coexistence. The average periodicity of the stripes, measured by fluorescence microscopy, is in the micrometer range, varying between 2 and 8 μm. The observed stripe-like patterns are stabilized due to dipole-dipole interactions between terminal $-CF_3$ groups. These interactions are particularly strong as compared with non-fluorinated lipids due to the low dielectric constant of the surrounding media (air). These long-range dipolar interactions tend to elongate the domains, in contrast to the line tension that tends to minimize the length of the domain boundary. This behavior should be compared with that of the lipid monolayer having alkyl chains, and which form spherical micro-domains (bubbles) at the gas-liquid coexistence. The measured stripe periodicity agrees quantitatively with a theoretical model. Moreover, the reduction in line tension by adding traces (0.1 mol %) of cholesterol results, as expected, in a decrease in the domain periodicity.

**Introduction**

Microscopic/mesoscopic "stripes" and "bubbles" in two and three dimensions can be found in a large variety of systems, including ferrofluids, [1] diblock copolymers, [2] noble gas adsorbed on surfaces, [3] and surfactant films. [4] These equilibrium patterns exhibit modulation in some order parameter (like polarization), and are stabilized by the interplay of competing intermolecular interactions. [5]

Domain shapes and shape transitions in lipid membranes have been a subject of intensive research for many years. The motivation to study lipid domains arises, for example, from the fundamental questions about the self-assembling process of micro-domains in plasma membranes, such as functional "rafts" of glycosphingolipids. [6] More specifically, insoluble Langmuir monolayers at the air/water interface have been studies in recent decades, where the thermodynamic variables such as surface concentration (area per molecule), lateral pressure and temperature can be controlled and measured using a Langmuir film balance. Furthermore, combination of film balance with microscopic techniques (e.g., florescence microscopy and Brewster angle microscopy) enables us to monitor domain size and shapes in the phase coexistence region. [7]

In the present study we focus on surfactants with partially fluoroalkylated hydrophobic tails (F-alkyl chains), which have been used for the design of new types of colloidal suspensions and versatile medical applications. [8-10] Due to the larger van der Waals radius of fluorine of about 1.35 Å, F-alkyl chains are known to form helices of six turns. [11] Several studies have been conducted to study the phase behavior of F-alkyl lipids at the air/water interface. For example, surface potential measurements of F-alkyl lipid monolayers in the liquid phase revealed that the molecular dipole of F-alkyl chains is stronger and points in the opposite direction than that of alkyl chains. [12,13] However, although the influence of molecular dipoles on the line tension is expected to be more evident for the phase coexistence at low surface pressures, a systematic characterization of such diluted phases is still missing.

In the present paper, we study the phase behavior of F-alkyl lipids anchored at air/water interface, by combining film balance experiments, fluorescence microscopy, and surface potential measurements. Surface potentials at very low surface pressures, i.e., in the gas-liquid coexistence region, could successfully be measured using a vibrating plate method. The shapes of liquid-phase domains were not found to be circular like for most alkyl lipids, but rather elongated, forming stripes of several micrometers in width. Such elongated domain shape as observed here can be theoretically attributed to the competition between electrostatic dipolar energy and line tension. This is discussed in the following sections.

**Materials and Methods**

**Materials.** The chemical structure of the F-alkyl and alkyl lipids is given in Fig. 1. The synthesis of F-alkyl lipid (FL-8-8) has been reported elsewhere.[14] Fluorescent tracers of DHPE (dihexadecyl phosphatidyl ethanolamine) labeled with Texas-Red from Molecular Probes Inc. (Leiden, Netherlands), was added for selective labeling of lipids in the liquid phase. De-ionized water (Millipore, Molsheim, France) was used as a subphase for monolayer experiments.

**Film Balance Experiments.** Langmuir isotherms and optical microscopic pictures are taken simultaneously. The surface pressure is measured using the Wilhelmy method coupled to a self-built rectangular Teflon trough with a total area of 47cm². The mixture of fluorinated lipids and fluorescence probes (molar ratio: 1000/1) is dissolved in a chloroform/methanol/water (65:25:4 v/v/v) solution, and deposited onto the subphase. After evaporation of the solvent (30 min), the film is compressed at a rate of approximately 1 Å²/s per molecule. The monolayer is illuminated with a monochromatic light ($\lambda$ = 581 nm from a mercury lamp. The fluorescence signal (at $\lambda$ = 601 nm) is collected with a 40 × LDW plan objective (Olympus, Hamburg, Germany) and a SIT camera (Hamamatsu, Herrsching,

Germany), and was recorded with a VCR. The images are then digitalized through open source software, NIH image (NIH, USA). Surface potentials are measured under well-defined thermodynamic conditions using the vibrating plate probe coupled to a µSpot mini-trough (Kibron OY, Helsinki, Finland).

**Results**

Fig. 2 shows Langmuir isotherms and fluorescence images of alkyl lipid (**AL-16**) in (a) and F-alkyl lipid (**FL-8-8**) in (b). The monolayers are kept at T = 20 ºC. As can be seen in the figure, **FL-8-8** occupies a larger area than **AL-16** at all surface pressures. For example, at $\Pi$ = 25 mN/m the area per **FL-8-8** molecule is 57 Å$^2$, while the corresponding value for **AL-16** is only 38 Å$^2$. These values are in good agreement with the cross sectional area twice that of a single F-alkyl chain, 30 Å$^2$, and that of a single alkyl chain, 19 Å$^2$, because there are two alkyl chains per lipid molecule. This is a consequence of the higher radius of the fluorinated groups in the lipid tails of FL-8-8 (Fig. 1), which leads to a helical conformation of the fluorcarbon chains as opposed to an all-trans (zigzag) conformation as known from studies of hydrocarbon chains. [11]

At very dilute monolayer states, above 100 Å$^2$ per molecule, no fluorescence emission can be detected from the Langmuir monolayer, indicating a gaseous phase (data not shown but see). [15] However, further compression of the film leads to the formation of bright domains (Fig. 2a and b). As presented in Fig. 2a alkyl lipids start to form circular bubble-like domains around 65 Å², whereas fluorinated lipid monolayers (Fig. 2b) form stripe-like domains appearing around 95 Å$^2$. Both domains can be attributed to the coexistence region of gas and liquid phases, because the dye-labeled lipid (DHPE Texas-Red) emits light only when surrounded by lipids in a liquid phase. This is further supported by the fact that homogenous fluorescence images (liquid phase) can be obtained by further compression to molecular areas below 50 for **AL-16 (2a)** or 85 Å² for **FL-8-8** (2b), respectively.

As presented in Fig. 2b, the width of the FL-8-8 stripe patterns varies between 2 - 8 µm. The area fraction of the liquid domains remains about 15%. Interestingly, the area per molecule seems to have only a minor influence on the width and on the area fraction in the phase coexistence region (between 97 and 89 Å²), up until a point near the onset of the pressure increase. There, all stripes disappear (85 Å²). It is well known that the transition from the gaseous to the liquid phase is difficult to determine from the pressure – area isotherm, [16]therefore no plateau in the isotherm can be observed.

It is worthwhile to mention that the images of the FL-8-8 monolayer appear homogeneously fluorescent over the entire area even at high surface pressures ($\Pi > 25$ mN/m). This suggests that the fluorescent lipids in a FL-8-8 monolayer remain in a fluid (disordered) environment even when the film reaches the minimum molecular area. The observed images are clearly different from a "dark" image of AL-16 taken at 25 mN/m (data not shown). This difference can be explained by the mismatch in the cross-sectional area of the ordered hydrocarbon (18 – 19 Å²) [17] and fluorocarbon chains (30 Å²). [11]Here, the alkyl chains of a FL-8-8 molecule are disordered due to the strain in the lateral chain packing, although the film is compressed to the area per molecule of 60 Å².

Furthermore, it was found that these stripe-like domains do not coalesce with each other, but rather maintain their elongated shape as individual domains. To eliminate kinetic effects arising from the finite compression speed, the FL-8-8 monolayer is kept for over 30 min at a constant area (90 Å² per molecule). Subsequent fluorescence images taken at 5 min intervals showed no changes in the global shape of the domains. In is possible to conclude that the observed stripe microdomains are thermodynamically stable. The F-Alkyl behavior can be compared with the image of the AL-16 monolayer in the vicinity of the gas-liquid coexistence (60 Å² per molecule). The latter exhibits circular fluorescent domains (Fig. 2a), as reported for other alkyl surfactants.[7]

Fig. 3 depicts the surface potential of a **FL-8-8** monolayer as a function of the area per fluorinated molecule, as well as the simultaneously measured Langmuir isotherm. In spite of some practical difficulties, as known from numerous previous studies,[18] we were able to determine the surface potential for diluted phases (larger than 90 Å$^2$ per molecule) and found values ranging between -420 mV and -390 mV. Increase in the surface pressure is highly correlated with an abrupt decrease in surface potential as is further addressed in the next section.

This finding demonstrates two remarkable features of F-alkyl lipids that are quite different from other lipids with alkyl chains: i) the surface potential remains negative for the entire range of measured surface areas and pressures; ii) compression of the film makes the surface potential even more negative. In fact, the corresponding surface potential of **AL-16** as a function of area per molecule is quite different from that of **FL-8-8**, but similar to many other alkyl lipids (data not shown).

**Discussion & Theoretical Modeling**

As confirmed by Fig. 2b, the observed stripe-like patterns are not caused by any kinetic effects, such as finite rate of film compression, because they are stable for a long times. The first-order phase transition between gas and liquid phases occurs as the attractive interaction between molecules becomes dominant against the entropy of mixing, similar to any condensation transition. The shape and thickness of coexisting domains is determined by two competing interactions.[19] The first is the line tension, namely, the energy needed to create a domain boundary between two phases at coexistence. This is proportional to the length of the line surrounding the domain and, therefore, tends to favor circular shapes, as observed for alkyl lipids. The second term is the molecular dipole moment, which tends to elongate (stretch) isolated domains due to their electrostatic interaction. Previous theoretical studies have actually postulated that the electrostatic repulsion in the gas-liquid phase coexistence stabilizes the phases with modulated density, by assuming the dipole contributions from polar head groups. [20,21]

Effective molecular dipole moments in Langmuir monolayers can be analytically estimated by surface potential measurements [22] using the Helmholtz equation,

$$V = \frac{\mu}{\varepsilon A} \tag{1}$$

where $\mu$ is the lipid dipole moment, $A$ the area per dipole, and $\varepsilon$ the media dielectric constant. To calculate the effective molecular dipole moment in Langmuir monolayers, three contributing factors are generally taken into account: [18] (i) contribution from the polar head group, (ii) influence of oriented (i.e. polarized) water molecules adjacent to the dipolar head group, and (iii) contribution from asymmetric chain termini. In our experiment, the intermediate chain regions do not contribute to the net molecular dipole because the successive dipoles along the linear chain cancel each other. As the contributions from (i) and (ii) are identical for F-alkyl (**FL-8-8**) and alkyl (**AL-16**) lipids,[23] the difference in surface potential seems to be dominated by the fluorinated chain termini. As found in previous studies [13,23], these chain termini have a dipole moment which points away from the monolayer towards the air. This is supported by the experimental observation of an increase in the absolute value of the (negative) surface potential as the film is compressed (fig. 3). This can be attributed to the lower surface area per molecule (see Eq. 1) and/or an in average more perpendicular orientation of the F-alkyl chains as the film is compressed.

If terminal $-CF_3$ dipoles reside in air, the electrostatic energy of the dipolar layer is given by

$$F_{el} = -\frac{1}{2} \int \frac{\vec{\mu} \cdot \vec{E}}{A} d^2r \tag{2}$$

where $\vec{E}$ is the electric field induced by the permanent dipoles and $A$ is the area per dipole.

The line tension $\gamma$ (in units of energy per unit length) accounts for the energy cost needed to create a domain of liquid phase in a gas background, on the two-dimensional monolayer. For an ideal, periodic arrangement of stripe-like domains with periodicity $D=D_L+D_G$, the contribution of the line tension to the domain energy is:

$$F_l = \frac{2\gamma}{D}. \qquad (3)$$

The electrostatic energy can be calculated within the assumption of an alternating arrangement of infinitely long stripes of liquid and gas domains having thickness $D_L$ and $D_G$, respectively. Using an additional assumption that the domain walls at the gas-liquid boundary are rather "sharp", the free energy for the stripe geometry is given by

$$F_{el} = \frac{kTb^3}{\pi a}\left[x\phi_L^2 + (1-x)\phi_G^2\right] - \frac{b^3}{\pi D}(\phi_L - \phi_G)^2 \log\left(\frac{D\sin(\pi x)}{\pi a}\right) \qquad (4)$$

where $x = D_L/D = D_L / (D_L+D_G)$ is the area fraction of the liquid phase, $a = (A)^{1/2}$ is a molecular length, $b^3 = \frac{\mu^2 \varepsilon}{kT\varepsilon_w(\varepsilon_w+\varepsilon)}$ and $kT$ is the thermal energy. Here, $\varepsilon$ and $\varepsilon_w$ represent the local dielectric constant near the dipoles and the dielectric constant of water, respectively, and $\phi_L$, $\phi_G$ are the concentrations (molecules per unit area) of the two phases. The first two terms in (4) represent the overall average contribution of the electrostatic energy, which is independent from the periodicity $D$. The third term is an exact summation of the dipole-dipole interaction. It contains both intra-stripe contributions proportional to $\log(D/a)/D$, and inter-stripe contributions proportional to $\log[\sin(\pi x)/\pi]/D$.[24] The total free energy difference between the stripe phase and a reference gas-liquid coexistence phase, therefore, is [25]

$$\Delta F = -\frac{kTb^3}{\pi D}(\phi_L - \phi_G)^2 \log\left(\frac{D\sin \pi x}{a\pi}\right) + \frac{2\gamma}{D}. \qquad (5)$$

By minimizing equation (5) with respect to the periodicity $D$, the equilibrium thickness of the stripes $D_L^{eq}$ can be obtained:

$$D_L^{eq} = a\left(\frac{\pi x}{\sin(\pi x)}\right)\exp\beta \qquad (6)$$

where

$$\beta = \frac{2\pi\gamma}{kTb^3(\phi_L - \phi_G)^2} + 1 \qquad (7)$$

By taking $\gamma = 1.6 \times 10^{-12}$ N, [26] $\varepsilon = 1$, and the measured surface potential into account, the equilibrium thickness of $D_L^{eq} \approx 1\mu m$ is obtained. This value is in good agreement with the experimental results presented in Fig. 2b (width of about 2 – 8 μm). As is confirmed experimentally, the assumption that phase coexistence reached thermodynamical is justified, and is clearly different from the nucleation and growth processes reported previously. [27] It should be noted that the choice of the dielectric constant value is critical due to the exponential dependence of $D$ on $\varepsilon$. To get the above estimate of $D_L^{eq}$, we chose the dielectric constant surrounding the -$CF_3$ dipoles to be that of air, $\varepsilon = 1$. If one takes $\varepsilon = 2$ (approximated dielectric constant of fluorine media), the resulting $D_L^{eq}$ increases by a factor of about 100. In fact, moving the dipoles from above to below the air/water interface, i.e. if one assumes that the dominant dipoles come from head groups in water, will result in a reduction of the interaction energy by a factor of 6400 due to the high dielectric constant of water ($\varepsilon_w = 80$). The negative surface potential, which is not found in regular alkyl lipids, indicates that the dipole moment originates mainly from the chain termini. Therefore, this is additional evidence that the dipolar major contribution comes from the chain residing above the water surface.

Equations (6) and (7) also suggest that a decrease in line tension $\gamma$ can cause a pronounced decrease in the periodicity $D$. The influence of line tension on $D$ is studied by adding a trace of cholesterol, which is known to reduce line tension significantly. [28] When 0.1 mol% of cholesterol is added to the fluorinated lipids, the average periodicity of stripe domains is decreased from 2 – 8 μm to 1 – 2 μm (Fig. 4). We note that it is experimentally difficult to reduce the doping ratio of cholesterol below 0.1 mol% or smaller. While doping of 0.2 mol% of cholesterol results in the complete disappearance of the stripe-like domains. However, the observed tendency supports the scenario postulated above, because equations (6) and (7) predict that the domain periodicity $D$ scales exponentially with the line tension $\gamma$.

**Conclusions**

We observed that surfactants with partially fluorinated lipid anchors (F-alkyl lipids, **FL-8-8**) form stripe-like domains at the gas-liquid phase coexistence. The domain periodicity ranged from 2 to 8 μm. Such stripe-like domains can be understood in terms of a competition between: (i) dipole-dipole interactions between chain termini that tend to elongate individual domains, and (ii) line tension that tends to minimize the length of the domain boundary. Because lipid monolayers with alkyl chains (**AL-16**) form spherical micro-domains (bubbles) at the gas-liquid coexistence, we conclude that the dipolar repulsion between $-CF_3$ termini is responsible for the domain formation and the resulting morphology. Surface potential measurement yields different sign for the potential of FL-8-8 monolayer as compared with regular alkyl chain and provides a further support for this conclusion. The stripe periodicity obtained by experiments is in agreement with the theoretical modeling. Moreover, as anticipated, a clear decrease in the stripe periodicity is observed when the line tension is reduced by trace (0.1 mol%) of cholesterol.

**Acknowledgement**

We thank E. Sackmann for helpful comments, C. Gege and R.R. Schmidt for synthesis of lipids, and P.K.J. Kinnunen for the surface potential measurements. This work was supported by Deutsche Forschungs Gemeinschaft (DFG Ta 259/2), Fonds der Chemischen Industrie, the Israel Science Foundation (ISF) under grant no. 210/01 and the US-Israel Binational Foundation (BSF) under grant no. 287/02. M.T. is thankful to DFG for a habilitation fellowship (Emmy Noether Program, Phase II), and D.A. thanks the Alexander von Humboldt Foundation for a research award.

**Figure Captions**

**Fig. 1** Chemical structures of (a) F-alkyl lipid (**FL-8-8**) and (b) alkyl lipid (**AL-16**) used as a reference. Note that only the lipid tails in (a) are fluorinated.

**Fig. 2** Langmuir isotherms of (a) **AL-16** and (b) **FL-8-8** monolayers at T = 20 ºC. In addition, fluorescence microscopy pictures are shown in (a) for 39, 43 and 61 Å², and in (b) for 85, 89, 92 and 97 Å². **AL-16** monolayers form spherical domains (bubbles), which are similar to those formed by other alkyl lipids. In contrast, **FL-8-8** monolayers at an area per molecule between 85 and 97 Å² form stripe-like micro-domains with an average stripe thickness ranging from 2 to 8 μm. These domains disappear near the onset of the surface pressure increase, corresponding to the area per molecule of 85 Å².

**Fig. 3** The surface potential of an FL-8-8 monolayer (right axis, solid line) plotted as a function of the area per F-alkyl molecule measured at 20°C. The isotherm with the same area per molecule is also given (broken line, left axis). The onset of the transition from the gas to liquid phase as observed by fluorescence microscopy is connected to an abrupt decrease in surface potential. An average surface potential close to the gas-liquid coexistence (approx. 90 Å² per molecule) was between -420 mV and -390 mV.

**Fig. 4** Fluorescence image of a FL-8-8 monolayer with 0.1 mol% cholesterol, taken at 90 A² per molecule (same area as in fig.2b) and 20°C. The doping of cholesterol leads to a decrease in the stripe periodicity from 2 – 8 μm to 1 – 2 μm.

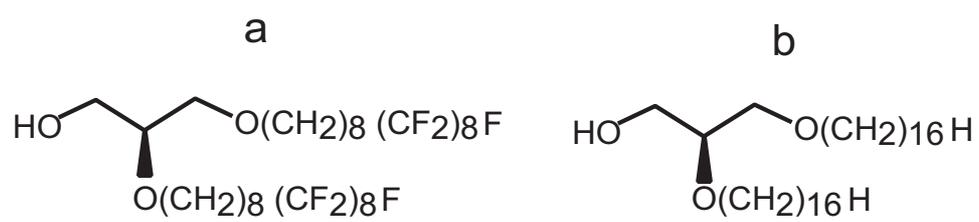

Schneider et al. Fig. 1

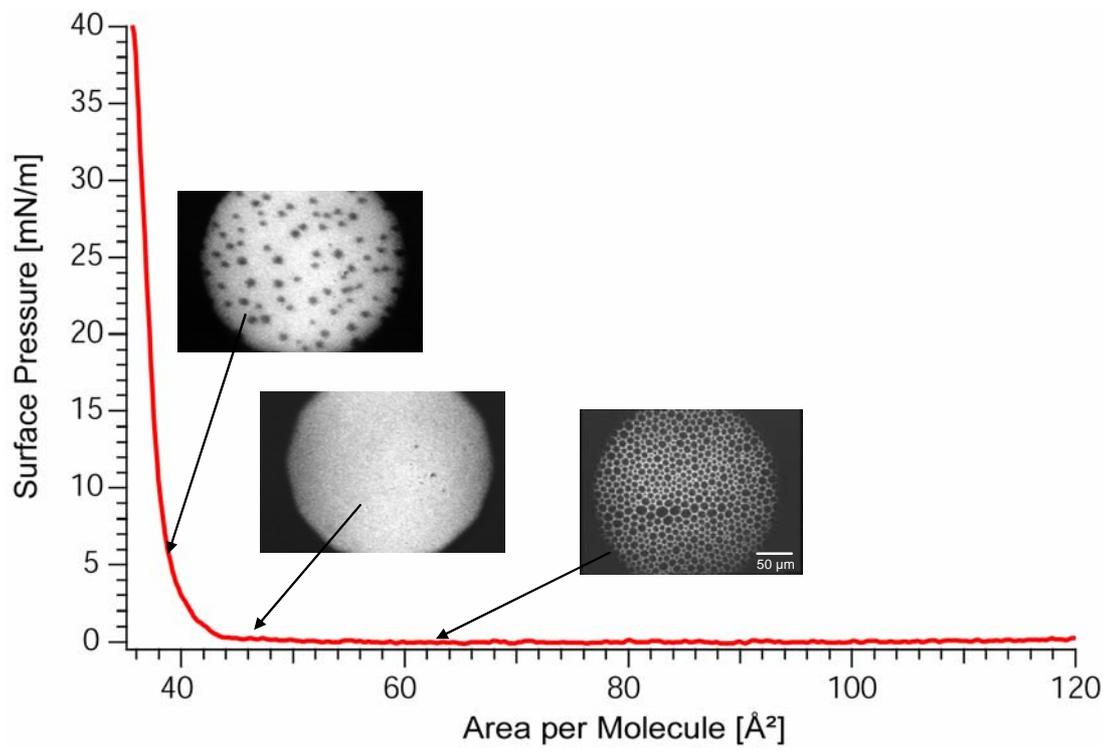

Schneider et al. Fig. 2a

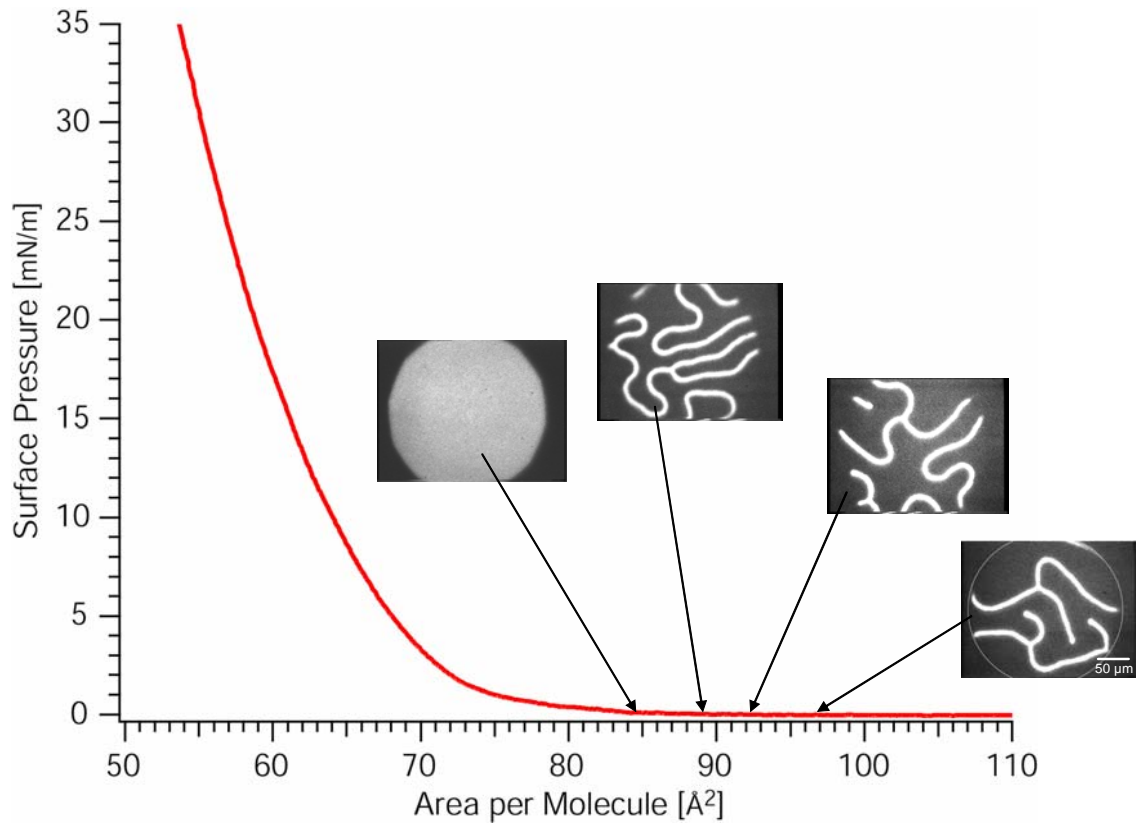

Schneider et al. Fig. 2b

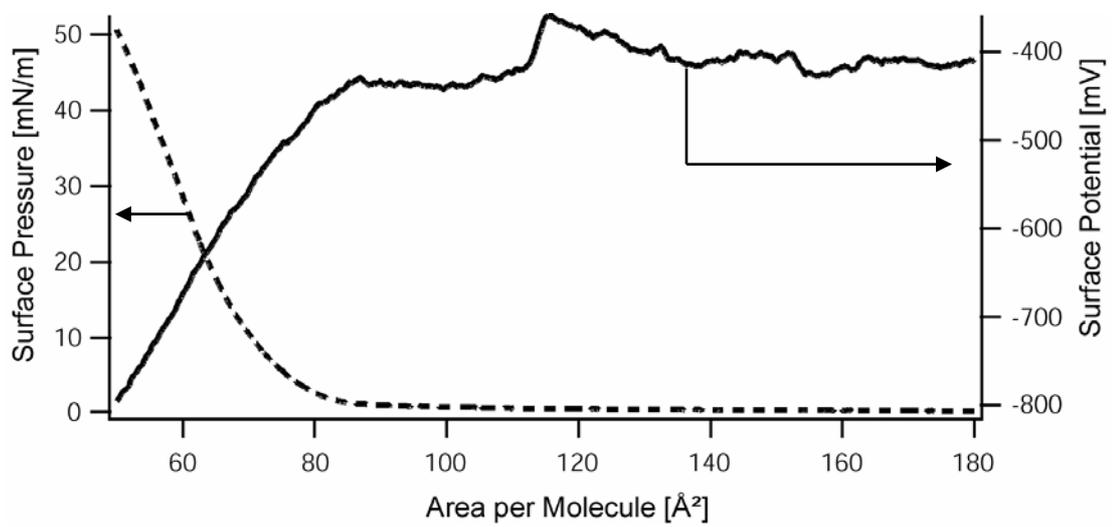

Schneider et al. Fig. 3

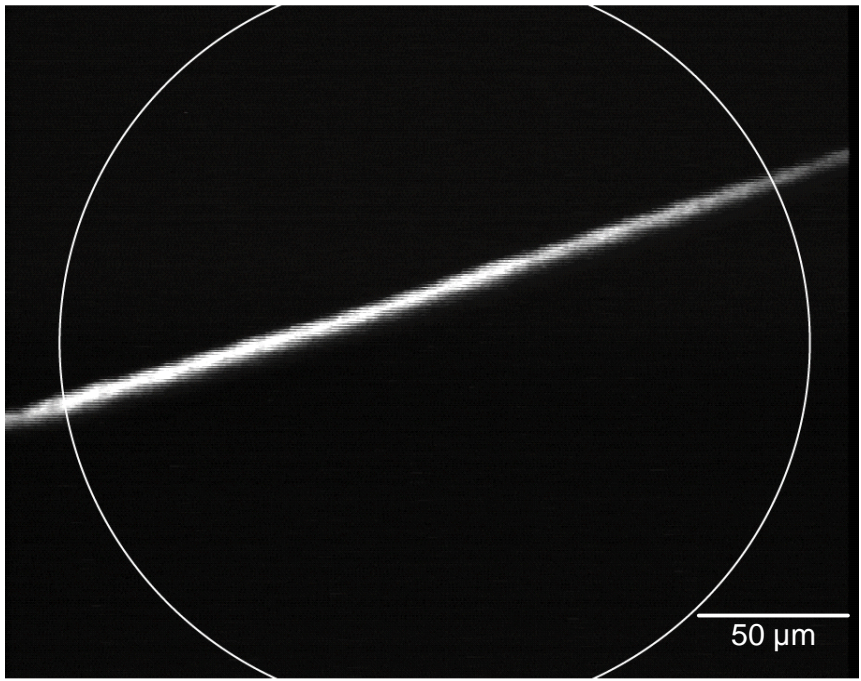

Schneider et al. Fig. 4